\documentclass[prl,twocolumn,aps,superscriptaddress,floatfix,nofootinbib]{revtex4-1}
\usepackage{amsmath}
\usepackage{amssymb}
\usepackage{simplewick}
\usepackage{graphicx}
\usepackage{color}
\newif\ifNoCount
\NoCounttrue 

\let\xxxhat\hat

\renewcommand{\hat}[1]{{\boldsymbol {\xxxhat {#1}} }}
\renewcommand{\vec}[1]{\boldsymbol {#1}}

\newcommand{\COMMENTED}[1]{}

\begin{document}
\title{Auxiliary Field quantum Monte Carlo for Strongly Paired Fermions}

\author{J. Carlson}
\author{Stefano Gandolfi}
\affiliation{
Theoretical Division, Los Alamos National Laboratory, Los Alamos, NM 87545, USA}
\COMMENTED{
\author{Shiwei Zhang}
\affiliation{
Department of Physics, College of William and Mary, Williamsburg, VA 23187, USA}
}
\author{Kevin E. Schmidt}
\affiliation{
Department of Physics, Arizona State University, Tempe, AZ 85287, USA}
\author{Shiwei Zhang}
\affiliation{
Department of Physics, College of William and Mary, Williamsburg, VA 23187, USA}

\begin{abstract}
We demonstrate that
the inclusion of a BCS importance function dramatically increases the
efficiency of the auxiliary field method for strong pairing.
We calculate the ground-state energy of an unpolarized fermi gas at
unitarity with up to 66 particles
and lattices of up to $27^3$ sites.
The method has no fermion sign problem, and an accurate result is obtained for the 
universal parameter $\xi$.
Several different forms of the kinetic
energy  adjusted to the unitary
limit but with different effective ranges extrapolate to the same continuum limit within error bars.
The  finite effective range results for different interactions are consistent with a linear
term proportional to the Fermi momentum times the effective range.
The new method described herein
will have many applications in superfluid
cold atom systems and in both electronic and nuclear
structures when pairing is important.
\end{abstract}

\date{\today} 


\ifNoCount
\maketitle 
\fi

The study of strongly interacting Fermi systems is one of the central themes and major challenges in physics. 
Superfluidity in unpolarized cold atomic Fermi gases, which has been demonstrated
both experimentally and theoretically, provides a prototypical example. The experimental ability to
use a Feshbach resonance to adjust the strength of the potential between
the atoms allows an exploration of the physics over many length scales.
A particularly interesting regime is at unitarity
where the scattering length diverges and the range of the potential
can be neglected. Since the particle density provides the only
length scale, the ground-state energy $E_0$ is proportional to the free fermi
gas energy $E_{FG}$,
\ifNoCount
\begin{equation}
\label{eq.xi}
E_0 = \xi E_{FG}\,. 
\end{equation}
\fi

The ability to quantitatively understand the properties of this system represents
a great triumph of many-body physics. 
Many experiments and calculations have been performed for the unitary Fermi
gas. Initial qualitative agreement was found between theory\cite{Carlson:2003,*Chang:2004,Astrakharchik:2004} and experiment\cite{Bartenstein:2004,Kinast:2005,Partridge:2006}.  More precise recent experiments have yielded $\xi = 0.39(2)$\cite{Luo:2009} and $0.41 (1)$\cite{Navon:2010}, with smaller values obtained very recently by Zwierlein, et al.\cite{Zwierlein:2011}.
Fixed-node Diffusion Monte Carlo (DMC)
calculations\cite{Carlson:2003,*Chang:2004,Astrakharchik:2004,
Carlson:2005,Lobo:2006,Forbes:2011,Gandolfi:2011}
have always included a
Bardeen-Cooper-Schrieffer\cite{Bardeen:1957} (BCS) trial
wave function to guide the Monte Carlo walk and provide the fixed
node constraint\cite{Anderson:1976} needed to overcome the fermion sign problem.
As is well known, these calculations provide an upper bound,
with the current best value
$\xi = 0.383(1)$\cite{Forbes:2011,Gandolfi:2011}.

In this paper we show that \emph{exact} calculations can be performed to 
accurately determine the ground-state properties of the unpolarized Fermi gas. 
A new method
is introduced to allow the use of a BCS trial wave function 
in the auxiliary-field quantum Monte Carlo (AFQMC) 
approaches of Zhang and
coworkers\cite{Zhang:1995,*Zhang:1997,Zhang:2003}. 
Using the new approach, we perform calculations
with several forms of the kinetic energy term that all give the correct
continuum limit but with different finite effective ranges to study the convergence with
particle number and lattice sizes, and to obtain the dependence of $\xi$ on the effective
range. An exact result is obtained for the value of $\xi$.

Quantum Monte Carlo simulations play a key role in addressing the challenge of 
strongly interacting Fermi systems.
The AFQMC method has been applied to a variety of systems in several fields.
With equal numbers and masses of up- and down-spin
fermions and an attractive interaction, there is no
fermion sign problem. 
The formalism presented here allows the use of a BCS importance function, which
drastically improves the efficiency in this situation.
In general applications, a sign or phase problem 
is present, which is controlled by a constraint, also using the importance function \cite{Zhang:1995,*Zhang:1997,Zhang:2003}.
Hartree-Fock or free Fermi gas (FG) type of importance functions have typically been used.
This approach
has been shown to be very accurate in many condensed matter models and 
optical lattices \cite{PhysRevLett.104.116402},
quantum chemistry \cite{al-saidi:224101}, and solid state materials \cite{PhysRevB.80.214116}.
\COMMENTED{
, for example, in molecules
comparable to the gold standard in quantum chemistry, the CCSD(T) method, and 
in solids often more accurate than fixed-node 
DMC with similar trial wave functions. 
}
Now BCS importance functions 
(or antisymmetrized geminal power (AGP) in chemistry)
will significantly improve our ability to deal with the sign problem in
systems where pairing is important, and enhance the capabilities for 
quantum simulations in strongly correlated systems in general.

The AFQMC method, in both-zero\cite{Lee:2006} and finite-temperature formulations\cite{Bulgac:2006}, has 
also been applied to the unitary Fermi gas.  Precise results require simultaneously
large lattices, so the system is dilute, and a large number of particles for an accurate 
approach to the continuum and thermodynamic limits.  Many such calculations have been
performed\cite{Lee:2006,Lee:2007,Lee:2008a,Lee:2008b,Bour:2011},
but the variance of the method limited the results to
relatively small number of particles and lattice sizes so that, as we demonstrate below,
the results are unlikely to have converged to the thermodynamic limit.

\COMMENTED{
Here we show how the auxiliary field methods developed by Zhang and
coworkers\cite{Zhang:1995,*Zhang:1997,Zhang:2003}
can be applied to this problem while including a BCS
trial function to guide the walk and greatly lower the variance.
We also perform calculations
with several kinetic energy terms that all give the correct
continuum limit to study the convergence with
particle number and lattice sizes.
}

The range of the van der Waals interaction in cold atoms is small
(e.g. about 3 nm in $^6$Li\cite{Ketterle:2008}) compared to interparticle
spacing so that the short range structure
of the interaction is unimportant; the results are completely
determined by the form of the kinetic energy and the scattering length.
For an $N_k^3$ lattice, the equivalent Hamiltonian is
\ifNoCount
\begin{equation}
\label{eq.latticeh}
H = \frac{1}{N_k^3} \sum_{\vec k,\vec j,\vec m, s}
\psi^\dagger_{\vec j s} \psi_{\vec m s} \epsilon_{\vec k}
e^{i \vec k\cdot (\vec r_{\vec j}-\vec r_{\vec m})}
+U \sum_{\vec i} n_{\vec i \uparrow} n_{\vec i\downarrow} \,.
\end{equation}
\fi
Here $\psi_{\vec js}$ is the destruction operator for a fermion of spin $s$ on
lattice site at position $\vec r_{\vec j}$. For odd lattice sizes
$N_k = 2N_c+1$, the $\vec k$ are given by
$\frac{2\pi}{L}(n_x \hat x+n_y\hat y+n_z\hat z)$ with
$-N_c \leq n_x,n_y,n_z \leq N_c$. The $k$ space destruction operators are
$c_{\vec k s} = N_k^{-3/2} \sum_{\vec j}
e^{-i\vec k \cdot \vec r_{\vec j}} \psi_{\vec js}$,
and the density operators are
$n_{\vec i s} = \psi^\dagger_{\vec i s}\psi_{\vec i s}$.

To reach the continuum limit, we need to take the limit of zero particle density, 
$\rho\equiv N/N_k^3\rightarrow 0$, in the context of the Hubbard model (i.e., replace $L$ by $N_k$).
Equivalently, because of scale invariance, we can think of the system as 
a discretized representation of a supercell with fixed size $L$, and
take the $k$-space cutoff to infinity. In either case, we then 
take the number of particles, $N$, to infinity.
\COMMENTED{
we would like to take the
$k$-space cutoff to infinity for fixed number of particles, $N$, and then take the
thermodynamic limit of the number of particles and the simulation cell
volume to infinity keeping their ratio constant.
}
In this limit only the behavior of $\epsilon_k$ for
$k \ll \frac{2\pi}{\alpha}$ is important, where $\alpha\equiv L/N_k$ is the lattice spacing. 
Thus a variety of kinetic energy
forms can be used as long as they are quadratic for $k$ much smaller than
the cutoff.
In this work we present results for
\ifNoCount
\begin{eqnarray}
\epsilon^{(2)}_k &=& \frac{\hbar^2 k^2}{2m}\,,
\ \ \epsilon^{(4)}_k = \frac{\hbar^2 k^2}{2m}
\left [ 1 -\beta^2 k^2\alpha^2 \right ]
\nonumber\\
\epsilon^{(h)}_{\vec k} &=&
\frac{\hbar^2}{m\alpha^2} \left [ 3-\cos (k_x\alpha)
-\cos (k_y\alpha) -\cos(k_z\alpha) \right ]
\,.
\end{eqnarray}
\fi
The superscript $2$ and $4$ indicate the highest power of $k$, while
$\epsilon^{(h)}$ is the Hubbard model hopping kinetic energy offset
by a constant so that it is zero at $\vec k=0$.

For two particles, the Hamiltonian is separable, and the solution of the
Lippmann-Schwinger equation for low-energy s-wave scattering gives
the phase-shift equation,
\ifNoCount
\begin{equation}
\begin{split}
&k \cot \delta_0 =
-\frac{4\pi\hbar^2}{mU\alpha^3}
\left [ \rule[-4mm]{1mm}{0mm}1 + 
\right .
\\
&\left . \frac{U \alpha^3}{16\pi^3} {\cal P}
\int_{-\pi/\alpha}^{\pi/\alpha} dk_x'
\int_{-\pi/\alpha}^{\pi/\alpha} dk_y'
\int_{-\pi/\alpha}^{\pi/\alpha} dk_z'
\frac{1}{\epsilon_{\vec k'}-\epsilon_{\vec k}} \right ]
\end{split}
\end{equation}
\fi
where ${\cal P}$ indicates the principal parts integration,
and the $k$ space sums are cut off by the lattice spacing $\alpha$.
The effective range expansion is
\ifNoCount
\begin{equation}
k \cot \delta_0 = -a^{-1} +\tfrac{1}{2} k^2 r_e + ...
\end{equation}
\fi
where $a$ is the scattering length and $r_e$ the effective range.
Since we are interested in the unitary limit, we adjust $U$ to have
$a^{-1}=0$. Both $\epsilon^{(2)}_k$ and $\epsilon^{(h)}_{\vec k}$ have
nonzero effective ranges. The extra parameter $\beta$ in $\epsilon^{(4)}_k$
can be adjusted to make the effective range zero.
The values for the parameters
are given in table \ref{t1}.
\begin{table}[h]
\ifNoCount
\begin{tabular}{|c|r|c|r|}
\hline
Energy & $U \tfrac{2m\alpha^2}{\hbar^2}$ & $\beta$ &
$r_e\alpha^{-1}$\\
\hline
$\epsilon^{(h)}_{\vec k}$ &  -7.91355 & -  & -0.30572\\
\hline
$\epsilon^{(2)}_k$ & -10.28871 & - &  0.33687 \\
\hline
$\epsilon^{(4)}_k$ & -8.66605  & 0.16137 & 0.00000\\
\hline
\end{tabular}
\fi
\caption{
The parameters that give infinite scattering length for two particles in an infinite lattice for the various kinetic energies. The $\beta$ value for the
$\epsilon^{(4)}_k$
kinetic energy has been adjusted to give zero effective
range, $r_e$.
}
\label{t1}
\end{table}

The AFQMC algorithm uses branching random walks to project the
ground state from an initial trial state with the 
imaginary-time operator $\exp [ - H \tau ]$.
Because the interaction is attractive, there is no fermion sign problem
for equal numbers of up and down fermions studied here, and no path constraint
is required. A walker is a set of $N$ single-particle orbitals.
Initially, the orbitals for the up spin particles are taken to be
identical to those for the down spin particles. The two-body
interaction term is broken up using a Hubbard-Stratonovich (HS) transformation
which has only positive weights, and treats the up and down spin particles
identically. Therefore, the up spin orbitals remain identical to the
down spin orbitals during the propagation. 
We will show below that the usual form for
a singlet paired BCS trial function also gives no fermion sign problem.

The walker states are given by specifying the orbital coefficients.
These can be specified on the real space lattice
$\phi_{n,\vec j}$ or as momentum space
coefficients $\tilde \phi_{n,\vec k}$
related to each other by a discrete Fourier transform. 
If we begin with real orbitals on the real space lattice, the orbitals
remain real when propagated. The momentum space orbitals therefore
satisfy
$\tilde \phi_{n, -\vec k} = \tilde \phi_{n, \vec k}^*$.
The orbitals are orthonormalized at each step. This is needed
to limit roundoff error, but the mathematical expressions are
correct without it.
The orthonormal orbitals therefore satisfy
$\sum_{\vec k} \tilde \phi_{n,\vec k}^* \tilde \phi_{m,\vec k} = \delta_{nm}$,
and the corresponding operators,
$w_{ns} = \sum_k \tilde \phi_{n,\vec k} c_{\vec ks}$,
satisfy
$\{w_{ns},w^\dagger_{ms'}\} = \delta_{nm}\delta_{ss'}$.
The walker state is
\ifNoCount
\begin{eqnarray}
|W\rangle &=&
\prod_{n=1}^{N/2} w^\dagger_{n\uparrow} w^\dagger_{n\downarrow}|0\rangle \,.
\end{eqnarray}
\fi
Because $N\ll N_k^3$ and because  the imaginary-time
history of the walk need not be retained, this formalism is much more efficient 
than the usual lattice formulations for the ground state of dilute gases.

Using a discrete HS 
transformation\cite{Hirsch:1983},
 the potential energy propagator is
 \ifNoCount
\begin{equation}
\begin{split}
&e^{-U\sum_i n_{\vec i \uparrow} n_{\vec i \downarrow}\Delta t} =
\frac{1}{2^{N_k^3}} \sum_{\{\sigma\} = \pm 1} 
G_V(\{\sigma\},\Delta t)\\
&G_V(\{\sigma\},\Delta t) =
\exp \left [ \sum_{\vec i,s} \left ( 2u\sigma_{\vec i}
-\tfrac{1}{2} U \Delta t \right ) n_{\vec i s}-u\sigma_{\vec i} \right ]\\
\end{split}
\end{equation}
\fi
where
$\tanh^2 u = -\tanh \left ( \frac{U \Delta t}{4} \right )$.
The kinetic energy propagator is
\ifNoCount
\begin{equation}
G_T(\Delta t) = \exp \left [ - \sum_{\vec k }
\epsilon_{\vec k} \left (
c^\dagger_{\vec k \uparrow}c_{\vec k \uparrow}
+c^\dagger_{\vec k \downarrow}c_{\vec k \downarrow} \right )\Delta t\right ] \,.
\end{equation}
\fi

Given a choice of one of the $N_k^3$ set of fields, the
application of the Trotter breakup
of one term of the propagator on
a walker $|W\rangle$ gives another walker $|W'\rangle$ times a weight 
$w'(\{\sigma\},W)$ that depends on the set of HS 
variables
$\{\sigma\}$ and $|W\rangle$,
\ifNoCount
\begin{eqnarray}
G(\{\sigma\},\Delta t)|W \rangle &\equiv& 
G_T(\tfrac{\Delta t}{2})
G_V(\{\sigma\},\Delta t)
G_T(\tfrac{\Delta t}{2})|W\rangle 
\nonumber \\
& \rightarrow & w'(\{\sigma\},W)|W'\rangle \,.
\end{eqnarray}
\fi
The propagation above consists of
(1) Multiply each $\tilde \phi_{n,\vec k}$ by
$\exp(-\tfrac{1}{2}\epsilon_{\vec k} \Delta t)$.
(2) Fast Fourier transform to obtain $\phi_{n,\vec i}$ in real space.
(3) Multiply each $\phi_{n,\vec i}$ by
$\exp\left ( 2u \sigma_i-\tfrac{1}{2} U\Delta t \right )$.
(4) Fast Fourier transform to obtain $\tilde \phi_{n,\vec k}$ in momentum space.
(5) Multiply each $\tilde \phi_{n,\vec k}$ by
$\exp(-\tfrac{1}{2}\epsilon_{\vec k} \Delta t)$. (6) Update the weight from non-operator terms.

Including importance sampling reduces the fluctuations, by changing
the sampling so that it is non-uniform, without biasing the results. We 
want to sample walkers $|W\rangle$ from
$\langle \Psi_T |W\rangle\langle W|\psi(t)\rangle$ where
\ifNoCount
\begin{equation}
|\psi(t+\Delta t)\rangle = e^{-(H-E_T)\Delta t} |\psi(t)\rangle
\end{equation}
\fi
The contribution of a walker $|W\rangle$ at the next time step is then
\ifNoCount
\begin{equation}
\label{eq.impsamp}
\begin{split}
&\sum_{\{\sigma\} = \pm 1} \left [
\frac{1}{2^{N_k^3}}
\frac{\langle \Psi_T| G(\{\sigma\},\Delta t) |W\rangle}{
\langle \Psi_T|W\rangle} e^{-E_T\Delta t} \right ] \times
\\
& \ \ \ \ \ 
\frac{1}{w(\{\sigma\},W)} G(\{\sigma\},\Delta t) |W\rangle \,.
\end{split}
\end{equation} 
\fi
We want to sample the set of HS 
variables $\{\sigma\}$,
from the unnormalized probability distribution given by the square
brackets. The normalization which, to order $\Delta t^2$ is
the local energy expression $e^{-(\tfrac{1}{2} [E_L(W)+E_L(W')]-E_T)\Delta t}$,
will give the weight of the sampled walkers. Once we have sampled
the $\{\sigma\}$ values, the new normalized walker is given by
the second term of Eq. \ref{eq.impsamp}.
We make sure that regions where the trial function is small are sampled 
adequately to eliminate trial-function bias.

%


The particle projected BCS state is
\ifNoCount
\begin{equation}
|BCS\rangle = \left [ \sum_{\vec k} f_k 
c^\dagger_{\vec k \uparrow} c^\dagger_{-\vec k\downarrow} \right ]^{N/2}
|0\rangle\,,
\end{equation}
\fi
where $f_k=v_k/u_k$ in the usual notation.
The overlap of a walker with the BCS state is
\ifNoCount
\begin{equation}
\langle W |BCS\rangle = \langle 0|
\prod_{n=1}^{N/2} w_{n\downarrow} w_{n\uparrow}
\left [ \sum_{\vec k} f_k   
c^\dagger_{\vec k \uparrow} c^\dagger_{-\vec k\downarrow} \right ]^{N/2}
|0\rangle
\end{equation}
\fi
The contraction needed is
$\contraction{}{w}{{}_{ns'}}{c}
w_{ns'}c^\dagger_{\vec k s} = \tilde \phi_{n,\vec k} \delta_{ss'}$.
The two creation operators in the BCS pair must be contracted with
some $w_{m\uparrow}$ and some $w_{n\downarrow}$, giving a term
\ifNoCount
\begin{equation}
\label{eq.bcselements}
\begin{split}
\contraction[2.5ex]{A_{nm}=}{w}{{}_{n\downarrow} w_{m\uparrow}
\sum_{\vec k} f_k 
c^\dagger_{\vec k\uparrow}}{c}
\contraction[2ex]{A_{nm}=w_{n\downarrow}}{w}{{}_{n\uparrow}
\sum_{\vec k} f_k 
}{c}
A_{nm}&=w_{n\downarrow} w_{m\uparrow}
\sum_{\vec k} f_k 
c^\dagger_{\vec k \uparrow} c^\dagger_{-\vec k\downarrow}
\\
&= \sum_{\vec k} \tilde \phi_{n,-\vec k} f_k 
\tilde \phi_{m,\vec k}
= \sum_{\vec k} \tilde \phi^*_{n,\vec k} f_k 
\tilde \phi_{m,\vec k}
\,.
\\
\end{split}
\end{equation}
\fi
Taking all the different possible full contractions,
\ifNoCount
\begin{equation}
\label{eq.overlap}
\langle W|BCS\rangle = {\rm det} A \,,
\end{equation}
\fi
where the elements of the $\tfrac{N}{2} \times \tfrac{N}{2}$
matrix $A$ are the $A_{nm}$ of Eq. \ref{eq.bcselements}.

The overlap, Eq. \ref{eq.overlap}, is positive when, as in the
standard singlet paired BCS solutions, $f_k\ge 0$. 
We write
${\rm det} A = {\rm det} \left [ B B^\dagger \right ]$
where $B^\dagger$ is the hermitian conjugate matrix of $B$ and
the matrix elements of the $\tfrac{N}{2} \times N_k^3$ matrix $B$ are
$B_{n\vec k} = \tilde \phi_{n,\vec k} \sqrt{f_k}$ \,.  
Applying the Cauchy-Binet theorem, each of the determinants of the submatrices
of $B$ is multiplied by the determinant of the corresponding hermitian
conjugate submatrix. The determinant of $A$ is a sum of positive terms,
so that our BCS trial function gives no sign problem.

Two estimates of the energy are used, the growth energy just measures the
growth of the weights in the random walk. The local energy can be calculated using contractions
similar to those above. Other observables can be calculated similarly. A complete
derivation for a general BCS state will be published elsewhere.
Here we give the result
\ifNoCount
\begin{equation}
\label{eq.locale}
\begin{split}
E_L(W) &= \frac{\langle W|H|BCS\rangle}{\langle W|BCS\rangle}
= 
2 {\rm tr} \left [ A^{-1} C \right ]
\\
&
+U\sum_{\vec q}  \left \{
{\rm tr} \left [ A^{-1} E(\vec q) \right ]
{\rm tr} \left [ A^{-1} E^\dagger(\vec q) \right ]
\right .
\\
&
\left .
-{\rm tr} \left [A^{-1} E(\vec q) A^{-1} E^\dagger (\vec q)\right ]
+{\rm tr} \left [ A^{-1} D(\vec q) \right ]
 \right \} \,,
\end{split}
\end{equation}
\fi
where
$D_{nm}(\vec q) =
\sum_{\vec k } \tilde \phi^*_{n,\vec k+\vec q} f_k 
\tilde \phi_{m,\vec k+\vec q}$,
$C_{nm}= \sum_{\vec k} \tilde\phi_{n,\vec k}^* \epsilon_{\vec k}
f_k \tilde \phi_{m,\vec k}$,
and
$E_{nm}(\vec q) =
\sum_{\vec k} \tilde \phi^*_{n,\vec k+\vec q} f_k \tilde 
\phi_{m,\vec k}$.
The
matrix elements of
$D$ and $E$ are convolutions
which are efficiently computed with fast
Fourier transforms. 
The computational cost of using the BCS $|\Psi_T\rangle$ is similar to using
a single Slater determinant.

\begin{figure}[th]
\ifNoCount
\includegraphics[width=.9\columnwidth]{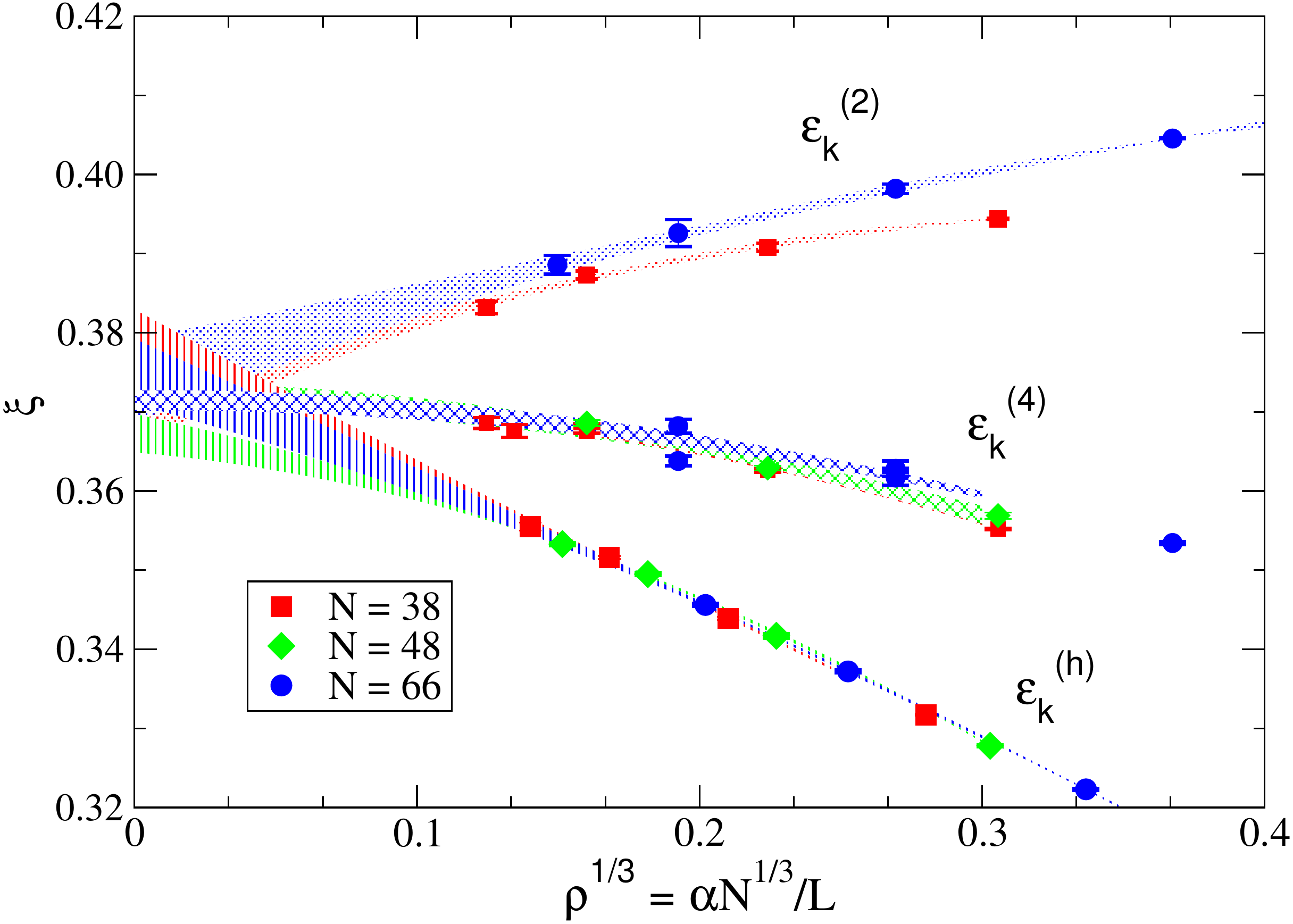}
\fi
\caption{(color online) The calculated ground state energy shown as the value of $\xi$
versus the lattice size for various particle numbers and Hamiltonians.
}
\label{fig.energy}
\end{figure}

We have calculated the ground-state energy for different particle numbers
and lattice sizes. The time-step errors have been extrapolated to zero
within statistical errors, and walker population biases have been checked
and were found to be negligible for the population sizes used. 
The imaginary time step is $\approx 0.01$\,-\,$0.05\  E_F^{-1}$ , the total
propagation time is of order $10$\,-\,$30\ E_F^{-1}$ and 
2,000-20,000 random walkers are used in the simulations. 

For $N=4$, we found that the use of BCS importance functions reduced the 
statistical error by a factor of $10$, or $100\times$ reduction in computer time, 
compared to the usual FG importance function.
The improvement increased to $1500\times$ for $N=38$ in a $12^3$ lattice. 
For larger systems, the discrepancy is much larger still; indeed the 
statistical fluctuations from the latter are such  
that often meaningful results cannot be obtained with the run configurations described above.

In Fig.~\ref{fig.energy} we summarize our calculations of the energy
as a function of $\rho^{1/3}$ where  $\rho=N/N_k^3$, and the particle
number is $N=38$, $48$ or $66$. We plot
$\xi$, Eq. \ref{eq.xi}, where we have in all cases used the infinite system
free-gas energy
$E_{FG} = \tfrac{3}{5} \tfrac{\hbar^2 k_F^2}{2 m}$ with
$k_F^3 =3 \pi^2 \frac{N}{\alpha N_k^3}$ as the reference.

\begin{table}[h]
\ifNoCount
\begin{tabular}{|l|r|c c|c c|}
\hline
Hamiltonian & $N$  & $\xi$ & err &  $A$ & err   \\
\hline
$\epsilon_k^{(2)}$ & 14 & 0.39 & 0.01 & 0.21 & 0.12 \\
            & 38 & 0.370 & 0.005 & 0.14 & 0.04 \\
            & 66 & 0.374 & 0.005 & 0.11 & 0.04 \\
\hline
$\epsilon_k^{(4)}$ & 38 & 0.372 & 0.002 &      &      \\
            & 48 & 0.372 & 0.003 &      &       \\
            & 66 & 0.372 & 0.003 &      &       \\
\hline
$\epsilon_k^{(h)}$ &  4 & 0.280 & 0.004 & -0.28& 0.05  \\
            & 38 & 0.380 & 0.005 & -0.17& 0.03  \\
            & 48 & 0.367 & 0.005 & -0.05& 0.03  \\
            & 66 & 0.375 & 0.005 & -0.13& 0.03  \\
\hline
\end{tabular}
\fi
\caption{ Energy extrapolations to infinite volume, zero range limit
for various particle numbers $N$ and different Hamiltonians. The term
linear in the effective range, $A$, is  also shown where it is not tuned to zero.
}
\end{table}

DMC calculations have found converged results when using
66 particles\cite{Forbes:2011,Gandolfi:2011}, and our results confirm this.
The differences between 38 and 66 particles are rather small. 
Our calculations with 14 particles show a significant size dependence, and with 26 particles 
the effects are still noticeable. These are not shown on the figure.
We have also computed the energy for 4 particle systems
for a variety of lattice sizes and find agreement with Ref. \cite{Bour:2011}.
The error bands in the figure provide least-squares estimates for the one
sigma error based upon quadratic fits to the finite-size effects.  The fits are
of the form $ E/ E_{FG} = \xi + A \rho^{1/3} + B \rho^{2/3}$.  For the interactions
tuned to $r_e = 0$, a fit with $A$ fixed to zero  is used.  Including a 
linear coefficient in the fit yields a value statistically consistent with zero.

The extrapolation in lattice size for the $k^2$ and Hubbard dispersions
show opposite slope as expected from the opposite signs of their
effective ranges.  
The extrapolation to $\rho \rightarrow 0$ is consistent with $\xi = 0.372 (0.005)$
in all cases.  Our final error contains statistical component and the errors associated
with finite population sizes and finite time-step errors. This value is below previous experiments, but more
compatible with recent experimental results of the Zwierlein group\cite{Zwierlein:2011}.

\begin{figure}[th]
\ifNoCount
\includegraphics[width=.9\columnwidth]{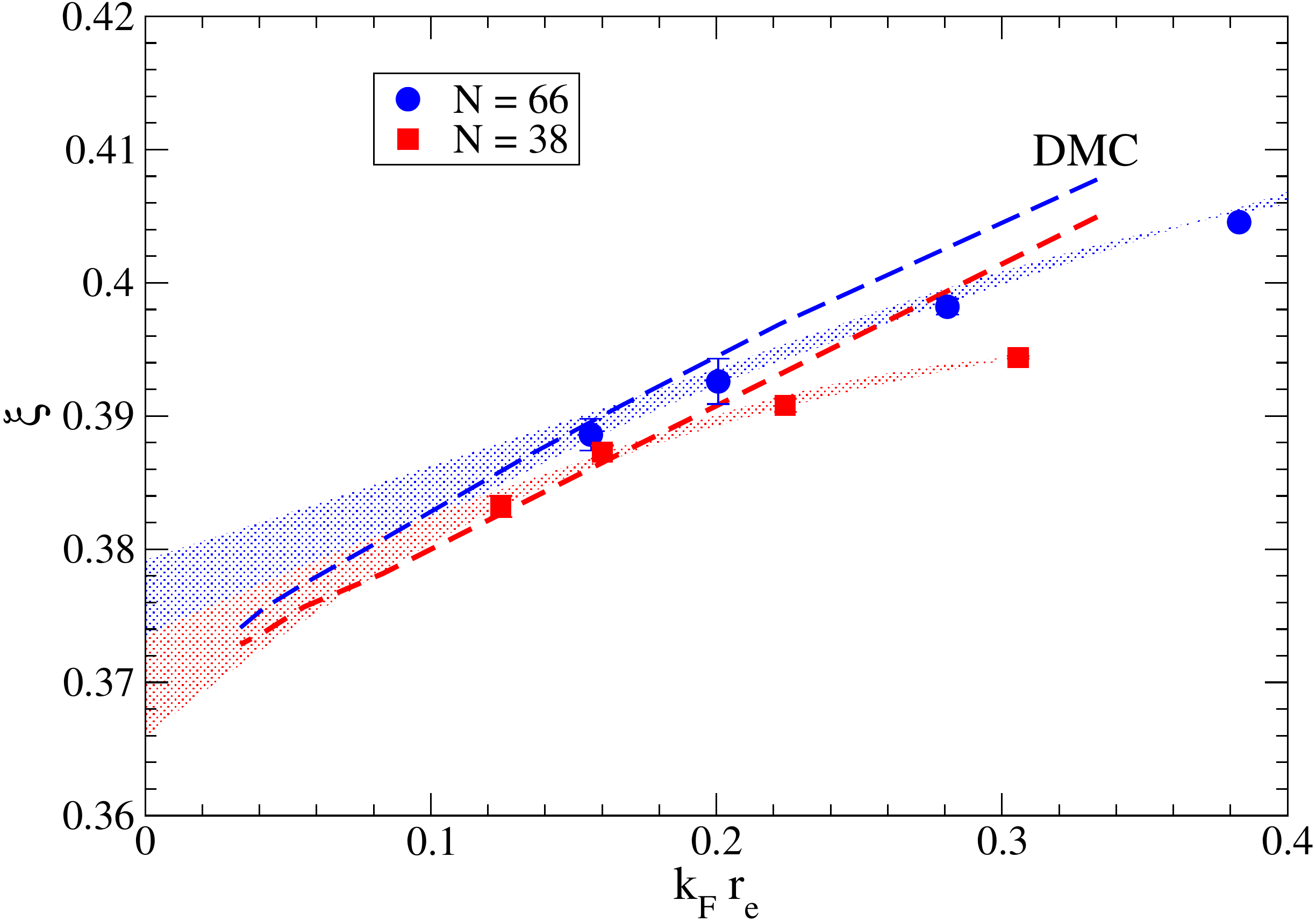}
\fi
\caption{(color online) The ground-state energy 
as a function of $k_F r_e$: 
comparison of DMC and AFQMC results. Dashed lines are DMC results, shifted down by 0.02 to enable comparison of the slopes.
}
\label{fig.range}
\end{figure}

We have also examined the behavior of the energy as a function of $k_F r_e$
for finite effective ranges. It has been conjectured\cite{Werner:2010} that the slope
of $\xi$ is universal: $\xi (r_e) = \xi + S k_F r_e$. Of course a finite range purely attractive interaction
is subject to collapse for a many-particle system, but a small repulsive many-body
interaction or the lattice,  
where double occupancy of a single species is not allowed, 
is enough to stabilize the system.   Our results are consistent with
the universality conjecture.  
In particular our results for zero effective range approach the continuum
limit with a slope consistent with zero. 

 Figure \ref{fig.range} compares the AFQMC results
for the $\epsilon_k^{(2)}$ interaction with the DMC results \cite{Forbes:2011,Gandolfi:2011} 
for various values of the effective range.  The AFQMC produces 
somewhat lower energies than the DMC, consistent with the upper-bound nature
of the DMC calculations. For the slope $S$ of $\xi$ with respect to finite $r_e$, the 
fit to the $N=66$ AFQMC results yields  $S = 0.11 (.03)  $.  Similar fits to the AFQMC data with the Hubbard dispersion $\epsilon_k^{(h)}$ for $N=66$ yield a linear term of $S = 0.12 (.03) $. Both are in agreement with the
DMC results of $S = 0.12 (.01)$.\cite{Gandolfia:2011}

In summary, we have shown how to incorporate a pairing importance function into auxiliary field quantum Monte Carlo
algorithms and used it to treat the unitary Fermi Gas.  This algorithm, for attractive interactions
and equal spin populations, is exact and can be extended to large lattices and strong
interactions. We find $\xi = 0.372 (.005)$ using a variety of interactions tuned to unitarity.
We also find a slope of the ground state energy with effective range of  S= 0.12 (.03) for the different
lattice and continuum Hamiltonians. This method should be useful without modification for
the entire BCS/BEC transition and for studying many other properties of cold Fermi gases.
It can also be applied to a wide variety of problems in other strongly-correlated fermions, in areas ranging from cold atoms to condensed matter to quantum chemistry to 
nuclear physics.

\ifNoCount
\begin{acknowledgments}
We thank Alexandros Gezerlis, Joaquin Drut and David B. Kaplan for useful discussions.
JC and SZ thank the Institute for Nuclear Theory (INT) at the University of Washington 
for its hospitality.
This work is supported by the U.S. Department of Energy,
Office of Nuclear Physics, under contracts DE-FC02-07ER41457
(UNEDF SciDAC), and DE-AC52-06NA25396 and by the National Science
Foundation grants PHY-0757703, PHY-1067777. KES thanks the Los Alamos National Laboratory
and the New Mexico Consortium for their hospitality. Computer time was also made
available by Los Alamos Open Supercomputing, NERSC, and CPD at William \& Mary.
We thank Chia-Chen Chang for help with computing and coding issues.
SZ is supported by ARO (56693-PH) and NSF (DMR-1006217).
\end{acknowledgments}


\begin{thebibliography}{10}%
\makeatletter
\providecommand \@ifxundefined [1]{%
 \ifx #1\undefined \expandafter \@firstoftwo
 \else \expandafter \@secondoftwo
\fi
}%
\providecommand \@ifnum [1]{%
 \ifnum #1\expandafter \@firstoftwo
 \else \expandafter \@secondoftwo
\fi
}%
\providecommand \enquote [1]{``#1''}%
\providecommand \bibnamefont  [1]{#1}%
\providecommand \bibfnamefont [1]{#1}%
\providecommand \citenamefont [1]{#1}%
\providecommand\href[0]{\@sanitize\@href}%
\providecommand\@href[1]{\endgroup\@@startlink{#1}\endgroup\@@href}%
\providecommand\@@href[1]{#1\@@endlink}%
\providecommand \@sanitize [0]{\begingroup\catcode`\&12\catcode`\#12\relax}%
\@ifxundefined \pdfoutput {\@firstoftwo}{%
 \@ifnum{\z@=\pdfoutput}{\@firstoftwo}{\@secondoftwo}%
}{%
 \providecommand\@@startlink[1]{\leavevmode}%
 \providecommand\@@endlink[0]{}%
}{%
 \providecommand\@@startlink[1]{%
  \leavevmode
  \pdfstartlink
   attr{/Border[0 0 1 ]/H/I/C[0 1 1]}%
   user{/Subtype/Link/A<</Type/Action/S/URI/URI(#1)>>}%
  \relax
 }%
 \providecommand\@@endlink[0]{\pdfendlink}%
}%
\providecommand \url  [0]{\begingroup\@sanitize \@url }%
\providecommand \@url [1]{\endgroup\@href {#1}{\urlprefix}}%
\providecommand \urlprefix [0]{URL }%
\providecommand \Eprint[0]{\href }%
\@ifxundefined \urlstyle {%
  \providecommand \doi [1]{doi:\discretionary{}{}{}#1}%
}{%
  \providecommand \doi [0]{doi:\discretionary{}{}{}\begingroup
  \urlstyle{rm}\Url }%
}%
\providecommand \doibase [0]{http://dx.doi.org/}%
\providecommand \Doi[1]{\href{\doibase#1}}%
\providecommand \bibAnnote [3]{%
  \BibitemShut{#1}%
  \begin{quotation}\noindent
    \textsc{Key:}\ #2\\\textsc{Annotation:}\ #3%
  \end{quotation}%
}%
\providecommand \bibAnnoteFile [2]{%
  \IfFileExists{#2}{\bibAnnote {#1} {#2} {\input{#2}}}{}%
}%
\providecommand \typeout [0]{\immediate \write \m@ne }%
\providecommand \selectlanguage [0]{\@gobble}%
\providecommand \bibinfo [0]{\@secondoftwo}%
\providecommand \bibfield [0]{\@secondoftwo}%
\providecommand \translation [1]{[#1]}%
\providecommand \BibitemOpen[0]{}%
\providecommand \bibitemStop [0]{}%
\providecommand \bibitemNoStop [0]{.\EOS\space}%
\providecommand \EOS [0]{\spacefactor3000\relax}%
\providecommand \BibitemShut [1]{\csname bibitem#1\endcsname}%
\bibitem{Carlson:2003}%
  \BibitemOpen
  \bibfield{author}{%
  \bibinfo {author} {\bibfnamefont{J.}~\bibnamefont{Carlson}}, \bibinfo
  {author} {\bibfnamefont{S.-Y.}\ \bibnamefont{Chang}}, \bibinfo {author}
  {\bibfnamefont{V.~R.}\ \bibnamefont{Pandharipande}},\ and\ \bibinfo {author}
  {\bibfnamefont{K.~E.}\ \bibnamefont{Schmidt}},\ }%
  \bibfield{journal}{%
  \Doi{10.1103/PhysRevLett.91.050401}{\bibinfo {journal} {Phys. Rev. Lett.}}\
  }%
  \textbf{\bibinfo {volume} {91}},\ \bibinfo {pages} {050401} (\bibinfo {year}
  {2003})%
  \bibAnnoteFile{NoStop}{Carlson:2003}%
\bibitem{Chang:2004}%
  \BibitemOpen
  \bibfield{author}{%
  \bibinfo {author} {\bibfnamefont{S.~Y.}\ \bibnamefont{Chang}}, \bibinfo
  {author} {\bibfnamefont{V.~R.}\ \bibnamefont{Pandharipande}}, \bibinfo
  {author} {\bibfnamefont{J.}~\bibnamefont{Carlson}},\ and\ \bibinfo {author}
  {\bibfnamefont{K.~E.}\ \bibnamefont{Schmidt}},\ }%
  \bibfield{journal}{%
  \Doi{10.1103/PhysRevA.70.043602}{\bibinfo {journal} {Phys. Rev. A}}\ }%
  \textbf{\bibinfo {volume} {70}},\ \bibinfo {pages} {043602} (\bibinfo {year}
  {2004})%
  \bibAnnoteFile{NoStop}{Chang:2004}%
\bibitem{Astrakharchik:2004}%
  \BibitemOpen
  \bibfield{author}{%
  \bibinfo {author} {\bibfnamefont{G.}~\bibnamefont{Astrakharchik}}, \bibinfo
  {author} {\bibfnamefont{J.}~\bibnamefont{Boronat}}, \bibinfo {author}
  {\bibfnamefont{J.}~\bibnamefont{Casulleras}},\ and\ \bibinfo {author}
  {\bibfnamefont{S.}~\bibnamefont{Giorgini}},\ }%
  \bibfield{journal}{%
  \Doi{10.1103/PhysRevLett.93.200404}{\bibinfo {journal} {Phys. Rev. Lett.}}\
  }%
  \textbf{\bibinfo {volume} {93}},\ \bibinfo {pages} {200404} (\bibinfo {year}
  {2004})%
  \bibAnnoteFile{NoStop}{Astrakharchik:2004}%
\bibitem{Bartenstein:2004}%
  \BibitemOpen
  \bibfield{author}{%
  \bibinfo {author} {\bibfnamefont{M.}~\bibnamefont{{Bartenstein}}}, \bibinfo
  {author} {\bibfnamefont{A.}~\bibnamefont{{Altmeyer}}}, \bibinfo {author}
  {\bibfnamefont{S.}~\bibnamefont{{Riedl}}}, \bibinfo {author}
  {\bibfnamefont{S.}~\bibnamefont{{Jochim}}}, \bibinfo {author}
  {\bibfnamefont{C.}~\bibnamefont{{Chin}}}, \bibinfo {author}
  {\bibfnamefont{J.~H.}\ \bibnamefont{{Denschlag}}},\ and\ \bibinfo {author}
  {\bibfnamefont{R.}~\bibnamefont{{Grimm}}},\ }%
  \bibfield{journal}{%
  \Doi{10.1103/PhysRevLett.92.203201}{\bibinfo {journal} {Phys. Rev. Lett.}}\
  }%
  \textbf{\bibinfo {volume} {92}},\ \bibinfo {pages} {203201} (\bibinfo {year}
  {2004})%
  \bibAnnoteFile{NoStop}{Bartenstein:2004}%
\bibitem{Kinast:2005}%
  \BibitemOpen
  \bibfield{author}{%
  \bibinfo {author} {\bibfnamefont{J.}~\bibnamefont{{Kinast}}}, \bibinfo
  {author} {\bibfnamefont{A.}~\bibnamefont{{Turlapov}}}, \bibinfo {author}
  {\bibfnamefont{J.~E.}\ \bibnamefont{{Thomas}}}, \bibinfo {author}
  {\bibfnamefont{Q.}~\bibnamefont{{Chen}}}, \bibinfo {author}
  {\bibfnamefont{J.}~\bibnamefont{{Stajic}}},\ and\ \bibinfo {author}
  {\bibfnamefont{K.}~\bibnamefont{{Levin}}},\ }%
  \bibfield{journal}{%
  \Doi{10.1126/science.1109220}{\bibinfo {journal} {Science}}\ }%
  \textbf{\bibinfo {volume} {307}},\ \bibinfo {pages} {1296} (\bibinfo {year}
  {2005})%
  \bibAnnoteFile{NoStop}{Kinast:2005}%
\bibitem{Partridge:2006}%
  \BibitemOpen
  \bibfield{author}{%
  \bibinfo {author} {\bibfnamefont{G.~B.}\ \bibnamefont{{Partridge}}}, \bibinfo
  {author} {\bibfnamefont{W.}~\bibnamefont{{Li}}}, \bibinfo {author}
  {\bibfnamefont{R.~I.}\ \bibnamefont{{Kamar}}}, \bibinfo {author}
  {\bibfnamefont{Y.-a.}\ \bibnamefont{{Liao}}},\ and\ \bibinfo {author}
  {\bibfnamefont{R.~G.}\ \bibnamefont{{Hulet}}},\ }%
  \bibfield{journal}{%
  \Doi{10.1126/science.1122876}{\bibinfo {journal} {Science}}\ }%
  \textbf{\bibinfo {volume} {311}},\ \bibinfo {pages} {503} (\bibinfo {year}
  {2006})%
  \bibAnnoteFile{NoStop}{Partridge:2006}%
\bibitem{Luo:2009}%
  \BibitemOpen
  \bibfield{author}{%
  \bibinfo {author} {\bibfnamefont{L.}~\bibnamefont{{Luo}}}\ and\ \bibinfo
  {author} {\bibfnamefont{J.~E.}\ \bibnamefont{{Thomas}}},\ }%
  \bibfield{journal}{%
  \Doi{10.1007/s10909-008-9850-2}{\bibinfo {journal} {J. Low Temp. Phys.}}\ }%
  \textbf{\bibinfo {volume} {154}},\ \bibinfo {pages} {1} (\bibinfo {year}
  {2009})%
  \bibAnnoteFile{NoStop}{Luo:2009}%
\bibitem{Navon:2010}%
  \BibitemOpen
  \bibfield{author}{%
  \bibinfo {author} {\bibfnamefont{N.}~\bibnamefont{Navon}}, \bibinfo {author}
  {\bibfnamefont{S.}~\bibnamefont{Nascimb{\`e}ne}}, \bibinfo {author}
  {\bibfnamefont{F.}~\bibnamefont{Chevy}},\ and\ \bibinfo {author}
  {\bibfnamefont{C.}~\bibnamefont{Salomon}},\ }%
  \bibfield{journal}{%
  \bibinfo {journal} {Science}\ }%
  \textbf{\bibinfo {volume} {328}},\ \bibinfo {pages} {729} (\bibinfo {year}
  {2010})%
  \bibAnnoteFile{NoStop}{Navon:2010}%
\bibitem{Zwierlein:2011}%
  \BibitemOpen
  \bibfield{author}{%
  \bibinfo {author} {\bibfnamefont{M.~W.}\ \bibnamefont{Zwierlein}}}%
   (\bibinfo {year} {2011}),\ \bibinfo {note} {talk presented at the Fermions
  from Cold Atoms to Neutron Stars Experimental Symposium, Institute for
  Nuclear Theory, Seattle, WA, May 16-20 2011}%
  \bibAnnoteFile{NoStop}{Zwierlein:2011}%
\bibitem{Carlson:2005}%
  \BibitemOpen
  \bibfield{author}{%
  \bibinfo {author} {\bibfnamefont{J.}~\bibnamefont{Carlson}}\ and\ \bibinfo
  {author} {\bibfnamefont{S.}~\bibnamefont{Reddy}},\ }%
  \bibfield{journal}{%
  \Doi{10.1103/PhysRevLett.95.060401}{\bibinfo {journal} {Phys. Rev. Lett.}}\
  }%
  \textbf{\bibinfo {volume} {95}},\ \bibinfo {pages} {060401} (\bibinfo {year}
  {2005})%
  \bibAnnoteFile{NoStop}{Carlson:2005}%
\bibitem{Lobo:2006}%
  \BibitemOpen
  \bibfield{author}{%
  \bibinfo {author} {\bibfnamefont{C.}~\bibnamefont{Lobo}}, \bibinfo {author}
  {\bibfnamefont{I.}~\bibnamefont{Carusotto}}, \bibinfo {author}
  {\bibfnamefont{S.}~\bibnamefont{Giorgini}}, \bibinfo {author}
  {\bibfnamefont{A.}~\bibnamefont{Recati}},\ and\ \bibinfo {author}
  {\bibfnamefont{S.}~\bibnamefont{Stringari}},\ }%
  \bibfield{journal}{%
  \Doi{10.1103/PhysRevLett.97.100405}{\bibinfo {journal} {Phys. Rev. Lett.}}\
  }%
  \textbf{\bibinfo {volume} {97}},\ \bibinfo {pages} {100405} (\bibinfo {year}
  {2006})%
  \bibAnnoteFile{NoStop}{Lobo:2006}%
\bibitem{Forbes:2011}%
  \BibitemOpen
  \bibfield{author}{%
  \bibinfo {author} {\bibfnamefont{M.~M.}\ \bibnamefont{Forbes}}, \bibinfo
  {author} {\bibfnamefont{S.}~\bibnamefont{Gandolfi}},\ and\ \bibinfo {author}
  {\bibfnamefont{A.}~\bibnamefont{Gezerlis}},\ }%
  \bibfield{journal}{%
  \Doi{10.1103/PhysRevLett.106.235303}{\bibinfo {journal} {Phys. Rev. Lett.}}\
  }%
  \textbf{\bibinfo {volume} {106}},\ \bibinfo {pages} {235303} (\bibinfo {year}
  {2011})%
  \bibAnnoteFile{NoStop}{Forbes:2011}%
\bibitem{Gandolfi:2011}%
  \BibitemOpen
  \bibfield{author}{%
  \bibinfo {author} {\bibfnamefont{S.}~\bibnamefont{Gandolfi}}, \bibinfo
  {author} {\bibfnamefont{K.~E.}\ \bibnamefont{Schmidt}},\ and\ \bibinfo
  {author} {\bibfnamefont{J.}~\bibnamefont{Carlson}},\ }%
  \bibfield{journal}{%
  \Doi{10.1103/PhysRevA.83.041601}{\bibinfo {journal} {Phys. Rev. A}}\ }%
  \textbf{\bibinfo {volume} {83}},\ \bibinfo {pages} {041601} (\bibinfo {year}
  {2011})%
  \bibAnnoteFile{NoStop}{Gandolfi:2011}%
\bibitem{Bardeen:1957}%
  \BibitemOpen
  \bibfield{author}{%
  \bibinfo {author} {\bibfnamefont{J.}~\bibnamefont{Bardeen}}, \bibinfo
  {author} {\bibfnamefont{L.~N.}\ \bibnamefont{Cooper}},\ and\ \bibinfo
  {author} {\bibfnamefont{J.~R.}\ \bibnamefont{Schrieffer}},\ }%
  \bibfield{journal}{%
  \Doi{10.1103/PhysRev.108.1175}{\bibinfo {journal} {Phys. Rev.}}\ }%
  \textbf{\bibinfo {volume} {108}},\ \bibinfo {pages} {1175} (\bibinfo {year}
  {1957})%
  \bibAnnoteFile{NoStop}{Bardeen:1957}%
\bibitem{Anderson:1976}%
  \BibitemOpen
  \bibfield{author}{%
  \bibinfo {author} {\bibfnamefont{J.~B.}\ \bibnamefont{Anderson}},\ }%
  \bibfield{journal}{%
  \bibinfo {journal} {J. Chem. Phys.}\ }%
  \textbf{\bibinfo {volume} {65}},\ \bibinfo {pages} {4121} (\bibinfo {year}
  {1976})%
  \bibAnnoteFile{NoStop}{Anderson:1976}%
\bibitem{Zhang:1995}%
  \BibitemOpen
  \bibfield{author}{%
  \bibinfo {author} {\bibfnamefont{S.}~\bibnamefont{Zhang}}, \bibinfo {author}
  {\bibfnamefont{J.}~\bibnamefont{Carlson}},\ and\ \bibinfo {author}
  {\bibfnamefont{J.~E.}\ \bibnamefont{Gubernatis}},\ }%
  \bibfield{journal}{%
  \Doi{10.1103/PhysRevLett.74.3652}{\bibinfo {journal} {Phys. Rev. Lett.}}\ }%
  \textbf{\bibinfo {volume} {74}},\ \bibinfo {pages} {3652} (\bibinfo {year}
  {1995})%
  \bibAnnoteFile{NoStop}{Zhang:1995}%
\bibitem{Zhang:1997}%
  \BibitemOpen
  \bibfield{author}{%
  \bibinfo {author} {\bibfnamefont{S.}~\bibnamefont{Zhang}}, \bibinfo {author}
  {\bibfnamefont{J.}~\bibnamefont{Carlson}},\ and\ \bibinfo {author}
  {\bibfnamefont{J.~E.}\ \bibnamefont{Gubernatis}},\ }%
  \bibfield{journal}{%
  \Doi{10.1103/PhysRevB.55.7464}{\bibinfo {journal} {Phys. Rev. B}}\ }%
  \textbf{\bibinfo {volume} {55}},\ \bibinfo {pages} {7464} (\bibinfo {year}
  {1997})%
  \bibAnnoteFile{NoStop}{Zhang:1997}%
\bibitem{Zhang:2003}%
  \BibitemOpen
  \bibfield{author}{%
  \bibinfo {author} {\bibfnamefont{S.}~\bibnamefont{Zhang}}\ and\ \bibinfo
  {author} {\bibfnamefont{H.}~\bibnamefont{Krakauer}},\ }%
  \bibfield{journal}{%
  \Doi{10.1103/PhysRevLett.90.136401}{\bibinfo {journal} {Phys. Rev. Lett.}}\
  }%
  \textbf{\bibinfo {volume} {90}},\ \bibinfo {pages} {136401} (\bibinfo {year}
  {2003})%
  \bibAnnoteFile{NoStop}{Zhang:2003}%
\bibitem{PhysRevLett.104.116402}%
  \BibitemOpen
  \bibfield{author}{%
  \bibinfo {author} {\bibfnamefont{C.-C.}\ \bibnamefont{Chang}}\ and\ \bibinfo
  {author} {\bibfnamefont{S.}~\bibnamefont{Zhang}},\ }%
  \bibfield{journal}{%
  \Doi{10.1103/PhysRevLett.104.116402}{\bibinfo {journal} {Phys. Rev. Lett.}}\
  }%
  \textbf{\bibinfo {volume} {104}},\ \bibinfo {pages} {116402} (\bibinfo {year}
  {2010})%
  \bibAnnoteFile{NoStop}{PhysRevLett.104.116402}%
\bibitem{al-saidi:224101}%
  \BibitemOpen
  \bibfield{author}{%
  \bibinfo {author} {\bibfnamefont{W.~A.}\ \bibnamefont{Al-Saidi}}, \bibinfo
  {author} {\bibfnamefont{S.}~\bibnamefont{Zhang}},\ and\ \bibinfo {author}
  {\bibfnamefont{H.}~\bibnamefont{Krakauer}},\ }%
  \bibfield{journal}{%
  \Doi{10.1063/1.2200885}{\bibinfo {journal} {J. Chem. Phys.}}\ }%
  \textbf{\bibinfo {volume} {124}},\ \bibinfo {eid} {224101} (\bibinfo {year}
  {2006})%
  \bibAnnoteFile{NoStop}{al-saidi:224101}%
\bibitem{PhysRevB.80.214116}%
  \BibitemOpen
  \bibfield{author}{%
  \bibinfo {author} {\bibfnamefont{W.}~\bibnamefont{Purwanto}}, \bibinfo
  {author} {\bibfnamefont{H.}~\bibnamefont{Krakauer}},\ and\ \bibinfo {author}
  {\bibfnamefont{S.}~\bibnamefont{Zhang}},\ }%
  \bibfield{journal}{%
  \Doi{10.1103/PhysRevB.80.214116}{\bibinfo {journal} {Phys. Rev. B}}\ }%
  \textbf{\bibinfo {volume} {80}},\ \bibinfo {pages} {214116} (\bibinfo {year}
  {2009})%
  \bibAnnoteFile{NoStop}{PhysRevB.80.214116}%
\bibitem{Lee:2006}%
  \BibitemOpen
  \bibfield{author}{%
  \bibinfo {author} {\bibfnamefont{D.}~\bibnamefont{Lee}},\ }%
  \bibfield{journal}{%
  \Doi{10.1103/PhysRevB.73.115112}{\bibinfo {journal} {Phys. Rev. B}}\ }%
  \textbf{\bibinfo {volume} {73}},\ \bibinfo {pages} {115112} (\bibinfo {year}
  {2006})%
  \bibAnnoteFile{NoStop}{Lee:2006}%
\bibitem{Bulgac:2006}%
  \BibitemOpen
  \bibfield{author}{%
  \bibinfo {author} {\bibfnamefont{A.}~\bibnamefont{{Bulgac}}}, \bibinfo
  {author} {\bibfnamefont{J.~E.}\ \bibnamefont{{Drut}}},\ and\ \bibinfo
  {author} {\bibfnamefont{P.}~\bibnamefont{{Magierski}}},\ }%
  \bibfield{journal}{%
  \Doi{10.1103/PhysRevLett.96.090404}{\bibinfo {journal} {Phys. Rev. Lett.}}\
  }%
  \textbf{\bibinfo {volume} {96}},\ \bibinfo {pages} {090404} (\bibinfo {year}
  {2006})%
  \bibAnnoteFile{NoStop}{Bulgac:2006}%
\bibitem{Lee:2007}%
  \BibitemOpen
  \bibfield{author}{%
  \bibinfo {author} {\bibfnamefont{D.}~\bibnamefont{Lee}},\ }%
  \bibfield{journal}{%
  \Doi{10.1103/PhysRevB.75.134502}{\bibinfo {journal} {Phys. Rev. B}}\ }%
  \textbf{\bibinfo {volume} {75}},\ \bibinfo {pages} {134502} (\bibinfo {year}
  {2007})%
  \bibAnnoteFile{NoStop}{Lee:2007}%
\bibitem{Lee:2008a}%
  \BibitemOpen
  \bibfield{author}{%
  \bibinfo {author} {\bibfnamefont{D.}~\bibnamefont{Lee}},\ }%
  \bibfield{journal}{%
  \Doi{10.1103/PhysRevC.78.024001}{\bibinfo {journal} {Phys. Rev. C}}\ }%
  \textbf{\bibinfo {volume} {78}},\ \bibinfo {pages} {024001} (\bibinfo {year}
  {2008})%
  \bibAnnoteFile{NoStop}{Lee:2008a}%
\bibitem{Lee:2008b}%
  \BibitemOpen
  \bibfield{author}{%
  \bibinfo {author} {\bibfnamefont{D.}~\bibnamefont{Lee}},\ }%
  \bibfield{journal}{%
  \bibinfo {journal} {Eur. Phys. J. A}\ }%
  \textbf{\bibinfo {volume} {35}},\ \bibinfo {pages} {171} (\bibinfo {year}
  {2008})%
  \bibAnnoteFile{NoStop}{Lee:2008b}%
\bibitem{Bour:2011}%
  \BibitemOpen
  \bibfield{author}{%
  \bibinfo {author} {\bibfnamefont{S.}~\bibnamefont{Bour}}, \bibinfo {author}
  {\bibfnamefont{X.}~\bibnamefont{Li}}, \bibinfo {author}
  {\bibfnamefont{D.}~\bibnamefont{Lee}}, \bibinfo {author}
  {\bibfnamefont{U.-G.}\ \bibnamefont{Meissner}},\ and\ \bibinfo {author}
  {\bibfnamefont{L.}~\bibnamefont{Mitas}},\ }%
  \bibfield{journal}{%
  \Doi{10.1103/PhysRevA.83.063619}{\bibinfo {journal} {Phys. Rev. A}}\ }%
  \textbf{\bibinfo {volume} {83}},\ \bibinfo {pages} {063619} (\bibinfo {year}
  {2011})%
  \bibAnnoteFile{NoStop}{Bour:2011}%
\bibitem{Ketterle:2008}%
  \BibitemOpen
  \bibfield{author}{%
  \bibinfo {author} {\bibfnamefont{W.}~\bibnamefont{Ketterle}}\ and\ \bibinfo
  {author} {\bibfnamefont{M.~W.}\ \bibnamefont{Zwierlein}},\ }%
  in\ \emph{\bibinfo {booktitle} {Ultracold Fermi Gases}},\ \bibinfo {editor}
  {edited by\ \bibinfo {editor} {\bibfnamefont{M.}~\bibnamefont{Inguscio}},
  \bibinfo {editor} {\bibfnamefont{W.}~\bibnamefont{Ketterle}},\ and\ \bibinfo
  {editor} {\bibfnamefont{C.}~\bibnamefont{Salomon}}}\ (\bibinfo {publisher}
  {IOSPress},\ \bibinfo {address} {Amsterdam},\ \bibinfo {year} {2008})\
  \Eprint{http://arxiv.org/abs/0801.2500v1}{arXiv:0801.2500v1}%
  \bibAnnoteFile{NoStop}{Ketterle:2008}%
\bibitem{Hirsch:1983}%
  \BibitemOpen
  \bibfield{author}{%
  \bibinfo {author} {\bibfnamefont{J.~E.}\ \bibnamefont{Hirsch}},\ }%
  \bibfield{journal}{%
  \Doi{10.1103/PhysRevB.28.4059}{\bibinfo {journal} {Phys. Rev. B}}\ }%
  \textbf{\bibinfo {volume} {28}},\ \bibinfo {pages} {4059} (\bibinfo {year}
  {1983})%
  \bibAnnoteFile{NoStop}{Hirsch:1983}%
\bibitem{Werner:2010}%
  \BibitemOpen
  \bibfield{author}{%
  \bibinfo {author} {\bibfnamefont{F.}~\bibnamefont{Werner}}\ and\ \bibinfo
  {author} {\bibfnamefont{Y.}~\bibnamefont{Castin}}}%
   (\bibinfo {year} {2010}),\
  \Eprint{http://arxiv.org/abs/1001.0774}{arXiv:1001.0774}%
  \bibAnnoteFile{NoStop}{Werner:2010}%
\bibitem{Gandolfia:2011}%
  \BibitemOpen
  \bibfield{author}{%
  \bibinfo {author} {\bibfnamefont{S.}~\bibnamefont{Gandolfi}}}%
   (\bibinfo {year} {2011}),\ \bibinfo {note} {fig. 2 includes updated DMC
  results at small $r_e$}%
  \bibAnnoteFile{NoStop}{Gandolfia:2011}%
\end{thebibliography}

%
\fi
\end{document}